\documentclass[12pt]{article}
\usepackage{amssymb}
\usepackage{graphicx}
\usepackage{indentfirst}
\usepackage{cite}

\begin{document}

\title{Temperature dependence of cross sections  \\
for meson-meson nonresonant reactions \\
in hadronic matter}
\author{Zhen-Yu Shen and Xiao-Ming Xu}
\date{}
\maketitle \vspace{-1cm}
\centerline{Department of Physics, Shanghai University,
Baoshan, Shanghai 200444, China}
\begin{abstract}
We study unpolarized cross sections for
the endothermic nonresonant reactions: $\pi\pi \to \rho\rho$ for $I=2$, 
$KK \to K^* K^*$ for $I=1$, $KK^* \to K^*K^*$ for $I=1$, 
$\pi K \to \rho K^*$ for $I=3/2$, $\pi K^* \to \rho K^*$ for $I=3/2$, 
$\rho K \to \rho K^*$ for $I=3/2$, and $\pi K^* \to \rho K$ for $I=3/2$, 
which take place in hadronic matter. We provide a
potential that is given by perturbative QCD with loop corrections at short
distances, becomes a distance-independent and temperature-dependent value
at long distances, and has a spin-spin interaction with relativistic
modifications. The Schr\"odinger equation with the potential yields
temperature-dependent meson masses and mesonic quark-antiquark relative-motion
wave functions. In the first Born approximation with the quark-interchange
mechanism, the temperature dependence of the potential, meson masses and wave
functions brings about temperature dependence of unpolarized cross sections
for the seven nonresonant reactions. Noticeably, rapid changes of $\pi$
and $K$ radii cause an increase in peak cross sections while the temperature
approaches the critical temperature. Parametrizations of the numerical cross
sections are given for their future applications in heavy ion collisions.
\end{abstract}


\newpage

\leftline{\bf 1. Introduction}
\vspace{0.5cm}

Heavy ion collisions at the Relativistic Heavy Ion
Collider (RHIC) and at the Large Hadron Collider (LHC) produce interesting
hadronic matter which final state can be detected. The simplest quantity
measured is the ratio of $p_T$-integrated midrapidity yields for mesons.
For central Au+Au collisions at $\sqrt{s_{NN}}$=200 GeV ratios from the
PHENIX Collaboration are $\pi^-/\pi^+=$0.984, $K^-/K^+=$0.933, 
$K^+/\pi^+=$0.171, and $K^-/\pi^-=$0.162 \cite{PHENIX}.
The STAR Collaboration obtained $\rho^{\rm 0}/\pi^-$=0.169 
for peripheral Au+Au collisions at $\sqrt{s_{NN}}$=200 GeV \cite{STAR1},
$\rho^{\rm 0}/\pi^-$=0.23 for midcentral Cu+Cu collisions at 
$\sqrt{s_{NN}}$=200 GeV \cite{STAR2}, and $K^{*\rm 0}/K^-$=0.2 for central
 Au+Au collisions at $\sqrt{s_{NN}}$=200 GeV \cite{STAR3}.
For central Pb+Pb collisions at $\sqrt{s_{NN}}=2.76$ TeV ratios from the ALICE
Collaboration are $\pi^-/\pi^+$=1, $K^-/K^+$=1, $K^+/\pi^+$=0.15, and
$K^-/\pi^-$=0.15 \cite{ALICE}.
The ratios indicate that pions, rhos, and kaons
are dominant species in hadronic matter. In the present work we concern
ourself about nonresonant reactions among $\pi$, $\rho$, $K$, 
and $K^*$ mesons, and focus on cross sections for the endothermic
nonresonant reactions: $\pi\pi \to \rho\rho$ for $I=2$, 
$KK \to K^* K^*$ for $I=1$, $KK^* \to K^*K^*$ for $I=1$, 
$\pi K \to \rho K^*$ for $I=3/2$,
$\pi K^* \to \rho K^*$ for $I=3/2$, 
$\rho K \to \rho K^*$ for $I=3/2$,
and $\pi K^* \to \rho K$ for $I=3/2$.
Cross sections for the inverse reactions of the seven reactions are obtained
by the detailed balance.

The mechanism that governs meson-meson nonresonant reactions is the
quark-inter- change process. With a choice of Gaussian quark-antiquark wave
functions the experimental data of $S$-wave elastic phase shifts for $\pi \pi$
scattering for $I=2$ and $K\pi$ scattering for $I=3/2$ 
in vacuum can be reproduced \cite{BS,BSJ}. Good quark-antiquark wave
functions are solutions of the Schr\"odinger equation with a quark potential
developed from QCD. Since nonperturbative QCD is not solved, the quark
potential is not unique. Meson-meson nonresonant reactions have been
studied in different quark potential models.
Dissociation cross sections of charmonia in collisions with $\pi$ and $\rho$
mesons were calculated in Refs. \cite{WSB,BSWX} with
the color Coulomb, spin-spin hyperfine, and linear confining interactions.
In a quark model with the potential that 
arises from one-gluon exchange plus perturbative one- and two-loop
corrections at the short distance and exhibits a linear form at the long
distance, we calculated unpolarized cross sections for the seven
nonresonant reactions in vacuum \cite{LX}. Hadronic matter attracts our
attention from the nonresonant reactions in vacuum to those in medium. 
One effect produced by high-temperature medium is that the linear
confinement at the long distance is replaced by a distance-independent
value which depends on temperature \cite{KLP}. In Ref. \cite{LX}
the spin-spin interaction contains the Fermi contact term and the term that
originates from the perturbative one- and two-loop corrections to 
the one-gluon exchange \cite{xmxu}. The Fermi contact term contains the
function $\delta^3 (\vec {r})$ which can
not be used in the Schr\"odinger equation. Then, quark-antiquark wave functions
are given by the Schr\"odinger equation
with only the central spin-independent potential, and 
$\pi$ and $\rho$ mesons ($K$ and $K^*$) have identical quark-antiquark
relative-motion wave functions. In the present work the function 
$\delta^3 (\vec {r})$ in the Fermi contact term is smeared to include
relativistic effects as done in Refs. \cite{GI,wong1} for meson spectra and
 in Refs. \cite{WSB,BSWX} for charmonium dissociation cross sections in vacuum.
This smearing 
significantly changes temperature dependence of $\pi$ and $K$ masses, and
causes different quark-antiquark relative-motion wave functions of $\pi$ and
$\rho$ mesons ($K$ and $K^*$). 
Hence, the cross sections for the seven nonresonant reactions are expected to
change significantly. Now we have arrived at a potential that is given by
perturbative QCD with loop corrections at the short distance, becomes a
distance-independent and temperature-dependent potential at the long distance,
and has the one-gluon-exchange spin-spin interaction smeared.
The potential leads to interesting temperature-dependent cross sections for
charmonium dissociation in collisions with pions and rhos in hadronic matter
\cite{ZX}. Therefore,
we must calculate the cross sections for the seven nonresonant reactions with
the quark potential. This is the subject of the present work.

This paper is organized as follows. In the next section we present
the potential of which the short-distance part is given by perturbative QCD
and of which the long-distance part is provided by the lattice calculations.
Temperature-dependent
meson masses and mesonic quark-antiquark relative-motion wave functions are
obtained from the Schr\"odinger equation with the potential. In Sec. 3
cross-section formulae are presented, numerical unpolarized cross
sections are displayed and parametrized, and relevant discussions are given.
Finally, we summarize the present work in Sec. 4.

\vspace{0.5cm}
\leftline{\bf 2. Meson masses}
\vspace{0.5cm}

The potential includes a central spin-independent potential and the smeared
spin-spin interaction. The central spin-independent potential is
\begin{equation}
V_{\rm {si}}(\vec {r})=-\frac {\vec {\lambda}_a}{2} \cdot \frac {\vec
{\lambda}_b}{2}
\frac{3}{4} D \left[ 1.3- \left( \frac {T}{T_{\rm c}} \right)^4 \right]
\tanh (Ar) 
+ \frac {\vec {\lambda}_a}{2} \cdot \frac {\vec {\lambda}_b}{2}
\frac {6\pi}{25} \frac {v(\lambda r)}{r} \exp (-Er),
\end{equation}
which shows the feature that the potential at long distances is a 
distance-independent value which decreases with increasing temperature. At
short distances no medium screening sets in, and $V_{\rm si}$ is the potential
arising from one-gluon exchange plus perturbative one- and two-loop
corrections. The loop corrections are a relativistic modification to the
one-gluon-exchange central spin-independent potential (the color Coulomb
interaction). In the potential 
$\vec r$ is the relative coordinate of constituents $a$ and $b$, and
$\lambda$=$\sqrt{\frac{3b_0}{16\pi^2 \alpha'}}$ with the Regge slope
$\alpha'$=1.04~Ge$V^{-2}$ and $b_0=11-\frac{2}{3}N_f$. In the case of four
quark flavors $N_f=4$, $A=1.5[0.75+0.25 (T/T_{\rm c} )^{10}]^6$ GeV,
$D=0.7$ GeV, and $E=0.6$ GeV. 
The function $v(x)$ is defined as \cite{BT}:
\[
v(x)=\frac
{4b_0}{\pi} \int^\infty_0 \frac {dQ}{Q} (\rho (\vec {Q}^2) -\frac {K}{\vec
{Q}^2}) \sin (\frac {Q}{\lambda}x),
\]
where the quantity $\rho (\vec {Q} ^2) -\frac {K}{\vec {Q} ^2}$ with
$K=\frac{3}{16\pi^2 \alpha'}$ is from the
one-gluon exchange plus the perturbative one- and two-loop corrections, and
$Q$ equals the absolute value of the gluon momentum $\vec Q$.

The smeared spin-spin interaction is \cite{ZX}
\begin{equation}
V_{\rm ss}(\vec{r})=
- \frac {\vec {\lambda}_a}{2} \cdot \frac {\vec {\lambda}_b}{2}
\frac {16\pi^2}{25}\frac{d^3}{\pi^{3/2}}\exp(-d^2r^2) 
\frac {\vec {s}_a \cdot \vec {s}_b} {m_am_b}
+ \frac {\vec {\lambda}_a}{2} \cdot \frac {\vec {\lambda}_b}{2}
  \frac {4\pi}{25} \frac {1} {r}
\frac {d^2v(\lambda r)}{dr^2} \frac {\vec {s}_a \cdot \vec {s}_b}{m_am_b},
\end{equation}
where $\vec {s}_a$ ($\vec {s}_b$) and $m_a$ ($m_b$) are the spin and mass of
constituent $a$ ($b$), respectively, and
\begin{equation}
d^2=\sigma_0^2[\frac{1}{2}+\frac{1}{2}(\frac{4m_a m_b}{(m_a+m_b)^2})^4]
+\sigma_1^2(\frac{2m_am_b}{m_a+m_b})^2,
\label{eq:d}
\end{equation}
with $\sigma_0=0.15$ GeV and $\sigma_1=0.705$.

The first term in the spin-spin interaction is a modification (smearing)
of the Fermi
contact term $- \frac {\vec {\lambda}_a}{2} \cdot \frac {\vec {\lambda}_b}{2}
\frac {16\pi^2}{25}\delta^3(\vec {r}) 
\frac {\vec {s}_a \cdot \vec {s}_b} {m_am_b}$ that comes from the one-gluon
exchange between constituents $a$ and $b$. The smearing
is achieved by the regularization $\delta^3 (\vec {r}) \to
\frac{d^3}{\pi^{3/2}}\exp(-d^2r^2)$. The use of the Fermi contact term in the
Schr\"odinger equation makes it impossible to solve the equation. However, the
smearing overcomes the difficulty and in addition incorporates relativistic
effects \cite{WSB,BSWX,GI,wong1}. The second term originates from the
perturbative one- and two-loop corrections to the one-gluon exchange. The loop
corrections are a relativistic modification to the one-gluon-exchange spin-spin
interaction. The potential 
$V_{\rm si}(\vec {r})+V_{\rm ss}(\vec {r})$ is a relativized potential.

The smeared spin-spin interaction given in Eq. (2) plus the central
spin-independent potential given in Eq. (1) is used in the Schr\"odinger 
equation to yield meson masses and mesonic quark-antiquark relative-motion wave
functions that differ from one another. 
Corresponding to the up and down quark masses $m_u=m_d=0.32$ GeV and the
strange quark mass $m_s=0.5$ GeV, the mass splittings of $\pi$, $\rho$, $K$, 
and $K^*$ involved in the present work are $m_{\rho}-m_{\pi}=0.6294$ GeV and
$m_{K^*}-m_K=0.39865$ GeV at $T=0$ in comparison with the experimental data of
0.6304 GeV and 0.3963 GeV. Temperature dependence of the $\pi$, $\rho$, $K$,
and $K^*$ masses is shown in Fig. 1 where the temperature covers the
temperature region of hadronic matter. While temperature increases,
the $\rho$ and $K^*$ masses decrease rapidly, and the $\pi$ and $K$ masses
decrease slowly for $0.6 \leq T/T_{\rm c} <0.88$ and rapidly for
$0.88 \leq T/T_{\rm c} <1$. The $\pi$ and $\rho$ mesons ($K$ and $K^*$)
become degenerate in mass at $T \to T_{\rm c}$. The $\pi$ and $\rho$ masses
approach zero while the temperature approaches the critical temperature. This
tendency is consistent with the dropping mass scenarios in hadronic matter in
Refs. \cite{weinberg,BLR}. According to the figure, the meson masses in
units of GeV in the region $0.6\leq T/T_{\rm c} < 1$ are parametrized as
\begin{eqnarray}
m_{\pi}&=&0.221 \left[ 1-\left( \frac{T}{0.99T_{\rm c}} \right)^{6.59}
\right]^{0.84},\nonumber \\
m_{\rho}&=&0.73 \left[ 1-\left( \frac{T}{0.992T_{\rm c}} \right)^{3.67}
\right]^{0.989},\nonumber \\
m_ {K} &=&0.46 \left[ 1-\left( \frac{T}{1.04T_{\rm c}} \right)^{8.58}
\right]^{0.88},\nonumber \\
m_ {K^*} &=&0.84 \left[ 1-\left( \frac{T}{1.05T_{\rm c}} \right)^{4.16}
\right].
\end{eqnarray}
The parametrization of the $\pi$ mass should be used to replace that given in 
Eq. (26) of Ref. \cite{ZX}, which is valid for $0.6 \leq T/T_{\rm c} < 0.97$.

In coordinate space the quark potential
\begin{equation}
V_{ab}(\vec{r})=V_{\rm {si}}(\vec{r})+V_{\rm {ss}}(\vec{r}),
\end{equation}
is used in the Schr\"odinger equation. But in some cases, for example, in
calculations of cross sections, the quark potential in momentum space is
convenient. The potential in momentum space is the Fourier transform of 
$V_{ab}(\vec{r})$:
\begin{eqnarray}
V_{ab}( \vec {Q} ) &=& - \frac{ \vec {\lambda } _ {a}} {2}
\cdot \frac{\vec {\lambda }_{b}}{2}\frac {3D}{4}
\left[ 1.3- \left( \frac {T}{T_{\rm c}} \right)^4 \right]
[(2\pi)^3\delta^3(\vec {Q})
-\frac {8\pi}{Q}\int^\infty_0 dr \frac {r\sin (Qr)}{\exp (2Ar)+1}]
\nonumber \\
&&+\frac{ \vec {\lambda }_{a}}{2} \cdot \frac{\vec {\lambda }_{b}}{2} 64 \pi E 
\int^\infty_0 dq \frac {\rho (q^2) -\frac{K}{q^2}}{(E^2+Q^2+q^2)^2-4Q^2q^2}  
\nonumber \\
&& -\frac{\vec {\lambda } _a}{2}\cdot \frac{\vec {\lambda }_{b}}{2}
\frac{16\pi ^{2}}{25}\exp \left( -\frac{Q^2}{4d^2} \right) 
\frac{\vec {s}_{a}\cdot \vec {s} _ {b}} {m_ {a} m_ {b}}
\nonumber \\
&&+\frac{\vec {\lambda }_{a}}{2}\cdot \frac{\vec {\lambda } _ {b}} {
2}\frac{16\pi ^{2}\lambda } {25Q} \int_{0}^{\infty}dx\frac{d^{2}v\left (
x\right) }{dx^{2}}\sin \left( \frac{Q}{\lambda }x\right)
\frac{\vec {s}_{a}\cdot \vec {s}_{b}}{m_{a}m_{b}}.
\end{eqnarray}

\vspace{0.5cm}
\leftline{\bf 3. Cross sections, numerical results and discussions}
\vspace{0.5cm}

Because of the temperature dependence of the quark potential, the meson masses
and mesonic quark-antiquark relative-motion wave functions are 
temperature-dependent. Now we are ready to use the temperature dependence of
the potential, masses and wave functions to obtain temperature
dependence of cross sections for the endothermic nonresonant reactions:
$\pi\pi \to \rho\rho$ for $I=2$, 
$KK \to K^* K^*$ for $I=1$, $KK^* \to K^*K^*$ for $I=1$, 
$\pi K \to \rho K^*$ for $I=3/2$,
$\pi K^* \to \rho K^*$ for $I=3/2$, 
$\rho K \to \rho K^*$ for $I=3/2$,
and $\pi K^* \to \rho K$ for $I=3/2$.

\vspace{0.5cm}
\leftline{\bf A. Cross-section formulae for meson-meson reactions}
\vspace{0.5cm}

When hadron-hadron scattering is studied, the center-of-mass motion of the two
hadrons needs to be separated to avoid frame-dependent results. 
One starts the study from a relativistic
Hamiltonian or a nonrelativistic Hamiltonian. It is impossible to separate off
the center-of-mass motion of the two scattering hadrons without taking any
approximation in the relativistic Hamiltonian, i.e., one can not change
the relativistic Hamiltonian \cite{QX} 
so that some terms contain the center-of-mass
coordinate and the other terms do not. We may keep
quark wave functions as Dirac spinors and the potential in a relativistic form,
and replace the relativistic kinetic energy of each quark by the 
nonrelativistic kinetic energy; the center-of-mass coordinate 
can then be separated off, and relativistic effects
still exist \cite{QX}. There is not unique relativistic formalism in treating
hadron-hadron scattering. Because nonrelativistic quark models have the
advantages: clear exhibition of hadron structures, easy formulation, effective
explanation of a good many experimental data, and useful applications in many
problems, as shown in Sec. 2, we establish the quark potential that is the
 one-gluon-exchange central spin-independent potential plus relativistic
 modifications, and use wave functions that are solutions of the Schr\"odinger
 equation with the relativized potential. The relativistic modifications are
 the perturbative one- and two-loop corrections to the gluon propagator
 \cite{xmxu,BT} and the smearing in the spin-spin interaction \cite{GI}.
 With the relativized potential the center-of-mass motion 
 is still exactly separated off. The center-of-mass system is thus chosen to
 conveniently study meson-meson scattering
$q_1\bar{q}_1 + q_2\bar{q}_2 \rightarrow q_1\bar{q}_2 + q_2\bar{q}_1$,
and cross sections depend on the momentum $\vec P$ of the initial meson 
$q_1\bar{q}_1$ and the momentum $\vec {P}^{\prime}$ of the final meson
$q_1\bar{q}_2$. $\vec P$ and $\vec {P}^{\prime}$ are related to the Mandelstam
variable $s=(P_{q_1\bar{q}_1 }+P_{q_2\bar{q}_2 } )^2$ by
\begin{equation}
\vec {P}^2(\sqrt{s})=\frac{1}{4s}\{ {[s-(
m_{q_1\bar {q}_1}^2+m_{q_2\bar {q}_2}^2)]^2
-4m_{q_1\bar {q}_1}^2m_{q_2\bar {q}_2}^2} \},
\end{equation}
\begin{equation}
\vec {P'}^{2}(\sqrt{s})=\frac{1}{4s} \{ {[
s- ( m_{q_1\bar {q}_2}^2+m_{q_2 \bar{q}_1}^2)]^2
-4m_{q_1 \bar {q}_2}^2m_{q_2 \bar{q}_1}^2} \},
\end{equation}
where $m_{q_1\bar {q}_1}$ and $P_{q_1\bar{q}_1}=(E_{q_1\bar{q}_1},
\vec{P}_{q_1\bar{q}_1})$ are the mass and the four-momentum of meson 
$q_1\bar {q}_1$, respectively. 
Similar notations are established for mesons $q_2\bar {q}_2$,
$q_1\bar {q}_2$ and $q_2\bar {q}_1$.

The cross section for $q_1\bar {q}_1 +q_2\bar {q}_2 \to q_1\bar {q}_2 
+q_2\bar {q}_1$ in the center-of-mass system is expressed as
\begin{equation}
\sigma(S,m_S,\sqrt {s},T) =\frac{1}{32\pi s}\frac{|\vec{P}^{\prime }(\sqrt{s})|
}{|\vec{P}(\sqrt{s})|}\int_{0}^{\pi }d\theta
|\mathcal{M}_{\rm fi} (s,t)|^{2}\sin \theta,
\label{eq:sigma}
\end{equation}
where $S$ is the total spin of either the two incoming mesons or the two
outgoing mesons, $m_S$ is the magnetic projection quantum number of $S$, 
$\mathcal{M}_{\rm fi}$ is the transition amplitude, $\theta$ is the angle 
between $\vec{P}$ and $\vec{P}'$, and the Mandelstam
variable $t=(P_{q_1\bar {q}_1}-P_{q_1\bar {q}_2})^2$.
The factor $\frac {1}{32\pi s} \frac {\mid \vec {P}' \mid}{\mid \vec {P} \mid}$
comes from the phase space element, the four-momentum conservation, and the 
relative flux of one initial meson with respect to another initial meson. While
$\sqrt s$ increases from the threshold energy of an endothermic reaction, 
the factor increases very rapidly and then decreases moderately.
The center-of-mass coordinate of mesons $q_1\bar {q}_1$ and $q_2\bar {q}_2$ 
($q_1\bar {q}_2$ and $q_2\bar {q}_1$) is not involved in the transition
amplitude, and the relative motion of constituents $q_1$, $\bar {q}_1$, $q_2$,
 and $\bar {q}_2$ makes a contribution
to the transition amplitude.

Meson-meson scattering process $q_1\bar{q}_1 +q_2\bar{q}_2 \to 
q_1\bar{q}_2 +q_2\bar{q}_1$ to lowest order includes one quark interchange and
one gluon exchange. The scattering has two forms: the prior form in which the
 gluon exchange takes place prior to the quark interchange and the post form
in which the quark interchange is followed by the gluon exhange.
The two forms of scattering may lead to different values of the transition
amplitude $\mathcal{M}_{\rm fi}$ \cite{MM,BBS,WC}. Hence, we denote the
transition amplitude in the prior form by $\mathcal{M}_{\rm fi}^{\rm prior}$
and the one in the post form by $\mathcal{M}_{\rm fi}^{\rm post}$. In
momentum space $\mathcal{M}_{\rm fi}^{\rm prior}$ and
$\mathcal{M}_{\rm fi}^{\rm post}$ are \cite{LX}
\begin{eqnarray}
&& {\cal M}_{\rm fi}^{\rm prior}=4\sqrt {E_{q_1 \bar q_1} E_{q_2 \bar {q}_2}
E_{q_1 \bar {q} _2} E_ {q_2 \bar {q} _1}}
\int\frac{d^3p_{q_1\bar{q}_2}}{(2\pi)^3}\frac{d^3p_{q_2\bar{q}_1}}{(2\pi)^3}
\nonumber   \\
&& \times\psi^\dagger_{q_1\bar {q}_2}(\vec{p}_{q_1\bar {q}_2})
\psi^\dagger_{q_2\bar {q}_1}(\vec{p}_{q_2\bar {q}_1})
(V_{q_1\bar {q}_2}+V_{\bar {q}_1q_2}+V_{q_1q_2}+V_{\bar {q}_1\bar {q}_2})
\psi_{q_1\bar {q}_1}(\vec{p}_{q_1\bar {q} _1}) \psi_{q_2\bar {q}_2}
(\vec{p}_{q_2\bar {q}_2}),
\end{eqnarray}
\begin{eqnarray}
&&{\cal M}_{\rm fi}^{\rm post}=4\sqrt {E_{q_1 \bar q_1} E_{q_2 \bar {q} _2} 
E_ {q_1 \bar {q} _2} E_ {q_2 \bar {q} _1}}           \nonumber\\
&&\times [\int\frac{d^3p_{q_1\bar{q}_1}}{(2\pi)^3}
\frac{d^3p_{q_1\bar{q}_2}}{(2\pi)^3}
\psi^\dagger_{q_1\bar {q}_2}(\vec{p}_{q_1\bar {q}_2})
\psi^\dagger_{q_2\bar {q}_1}(\vec{p}_{q_2\bar {q}_1}) V_{q_1\bar {q} _1} 
\psi_{q_1\bar {q}_1}(\vec{p}_{q_1\bar {q}_1}) \psi_{q_2\bar {q}_2}
(\vec{p}_{q_2\bar {q}_2})              \nonumber\\
&&+\int\frac{d^3p_{q_2\bar{q}_2}}{(2\pi)^3}\frac{d^3p_{q_2\bar{q}_1}}{(2\pi)^3}
\psi^\dagger_{q_1\bar {q}_2}(\vec{p}_{q_1\bar {q}_2})
\psi^\dagger_{q_2\bar {q}_1}(\vec{p}_{q_2\bar {q}_1}) V_{\bar {q}_2 q_2}
\psi_{q_1\bar {q}_1}(\vec{p}_{q_1\bar {q}_1}) \psi_{q_2\bar {q}_2}
(\vec{p}_{q_2\bar {q}_2})              \nonumber\\
&&+\int\frac{d^3p_{q_1\bar{q}_2}}{(2\pi)^3}\frac{d^3p_{q_2\bar{q}_1}}{(2\pi)^3}
\psi^\dagger_{q_1\bar {q}_2}(\vec{p}_{q_1\bar {q}_2})
\psi^\dagger_{q_2\bar {q}_1}(\vec{p}_{q_2\bar {q}_1}) V_{q_1q_2} 
\psi_{q_1\bar {q}_1}(\vec{p}_{q_1\bar {q}_1}) \psi_{q_2\bar {q}_2}
(\vec{p}_{q_2\bar {q}_2})              \nonumber\\
&&+\int\frac{d^3p_{q_1\bar{q}_2}}{(2\pi)^3}\frac{d^3p_{q_2\bar{q}_1}}{(2\pi)^3}
\psi^\dagger_{q_1\bar {q}_2}(\vec{p}_{q_1\bar {q}_2})
\psi^\dagger_{q_2\bar {q}_1}(\vec{p}_{q_2\bar {q}_1}) V_{\bar {q}_1\bar {q}_2} 
\psi_{q_1\bar {q}_1}(\vec{p}_{q_1\bar {q} _1}) 
\psi_{q_2\bar {q}_2}(\vec{p}_{q_2\bar {q} _2}) ],
\end{eqnarray}
where $\psi_{ab}(\vec{p}_{ab})$ is the product of color, spin, flavor and
relative-motion wave functions of constituents $a$ and $b$, and satisfies
$\int \frac {d^3p_{ab}}{(2\pi)^3} \psi^+_{ab} 
(\vec {p}_{ab})\psi_{ab} (\vec {p}_{ab}) =1,$
where $\vec{p}_{ab}$ is the relative momentum of $a$ and $b$. 

Cross sections corresponding to the scattering in the prior form and in the 
post form are
\begin{equation}
\sigma^{\rm prior}(S,m_S,\sqrt {s},T)
=\frac{1}{32\pi s}\frac{|\vec{P}^{\prime }(\sqrt{s}) |
}{|\vec{P}(\sqrt{s})|}\int_{0}^{\pi }d\theta
|\mathcal{M}_{\rm fi} ^ {\rm prior} (s,t)|^{2}\sin \theta ,
\label{eq:sigma prior}
\end{equation}
and
\begin{equation}
\sigma^{\rm post}(S,m_S,\sqrt {s},T)
=\frac{1}{32\pi s}\frac{|\vec{P}^{\prime }(\sqrt{s}) |
}{|\vec{P}(\sqrt{s})|}\int_{0}^{\pi }d\theta
|\mathcal{M}_{\rm fi} ^ {\rm post} (s,t)|^{2}\sin \theta ,
\label{eq:sigma post}
\end{equation}
respectively. The unpolarized cross section for 
$q_1\bar {q} _1 + q_2 \bar {q} _2 \to 
q_1 \bar {q} _2 + q_2 \bar {q} _1$ is
\begin{eqnarray}
\sigma^{\rm unpol} (\sqrt {s}, T) &=&
\frac{1}{(2S_{q_1 \bar q_1} +1) (2S_ {q_2 \bar q_2}+1)}\sum\limits_S (2S+1)
\nonumber \\
&&\times\frac{\sigma^{\rm prior}(S,m_S,\sqrt {s},T)
+\sigma^{\rm post}(S,m_S,\sqrt {s},T)}{2},  
\label{eq:sigma unpol}
\end{eqnarray}
where $S_{q_1 \bar q_1}$ and $S_{q_2 \bar q_2}$ are the spins of 
$q_1 \bar q_1$ and $q_2\bar q_2$, respectively. Even though the 
unpolarized cross section
is calculated in the center-of-mass system, its frame independence and
its dependence on the Mandelstam
variable ensure that it can be used in any frame.

\vspace{0.5cm}
\centerline{\bf B. Numerical results}
\vspace{0.5cm}

In Figs. 2-8 we plot unpolarized cross sections for the following
nonresonant reactions: $\pi\pi \to \rho\rho$ for $I=2$, 
$KK \to K^* K^*$ for $I=1$, $KK^* \to K^* K^*$ for $I=1$, 
$\pi K \to \rho K^*$ for $I=3/2$, $\pi K^* \to \rho K^*$ for $I=3/2$, 
$\rho K \to \rho K^*$ for $I=3/2$, and $\pi K^* \to \rho K$ for $I=3/2$.
These reactions are endothermic, and the cross section for each reaction
at a given temperature has at least one peak near the threshold energy
$\sqrt {s_0}$. We use a function of the form
$a \left( \frac {\sqrt {s} -
\sqrt {s_0}} {b} \right)^{c}\exp \left[ c \left( 1-\frac {\sqrt {s} -
\sqrt {s_0}} {b} \right) \right]$ to fit the numerical cross section. The
parameters $a$, $b$, and $c$ are adjusted to get the height, the
width of the peak, and the peak's location on the $\sqrt s$-axis. The
function with one peak can well fit some curves of which each has a peak,
but is completely not enough to fit a curve with two peaks. To remedy this,
we use a sum of two functions
\begin{small}
\begin{eqnarray}
\sigma^{\rm unpol}(\sqrt {s},T) & = &
a_1 \left( \frac {\sqrt {s} -\sqrt {s_0}} {b_1} \right)^{c_1}
\exp \left[ c_1 \left( 1-\frac {\sqrt {s} -\sqrt {s_0}} {b_1} \right) \right]
\nonumber \\
& + & a_2 \left( \frac {\sqrt {s} -\sqrt {s_0}} {b_2} \right)^{c_2}
\exp \left[ c_2 \left( 1-\frac {\sqrt {s} -\sqrt {s_0}} {b_2} \right) \right],
\end{eqnarray}
\end{small}
to get a satisfactory fit. This parametrization contains two terms. No more
terms are used because of terrible computation time. Parameters' values are
listed in Tables 1 and 2. We have calculated unpolarized cross sections only
at the six temperatures $T/T_{\rm c}=0$, 0.65, 0.75, 0.85, 0.9, 0.95. 
Unpolarized cross sections at any temperature that is between
$0.65T_{\rm c}$ and $T_{\rm c}$ can be obtained via the linear interpolation or
 extrapolation among the
cross sections at the five temperatures $0.65T_{\rm c}$, $0.75T_{\rm c}$, 
$0.85T_{\rm c}$, $0.9T_{\rm c}$, and $0.95T_{\rm c}$. A procedure to achieve
this can be found in Ref. \cite{ZX}. The procedure needs the quantity
 $d_{\rm 0}$ what is difference of the peak's location on the
 $\sqrt s$-axis with respect to the threshold energy and the quantity
 $\sqrt {s_{\rm z}}$ what is the quare root of the Mandelstam variable at which
 the cross section is 1/100 of the peak cross section. The two quantities are
 listed in Tables 1 and 2.

\vspace{0.5cm}
\centerline{\bf C. Discussions}
\vspace{0.5cm}

While $\sqrt s$ increases, $\int_0^\pi d\theta \mid {\cal M}_{\rm fi}(s,t) 
\mid^2\sin \theta$ in the cross-section formula decreases slowly near the 
threshold energy and rapidly in the other region, and by contrast the factor
$\frac {1}{s}\frac {\mid \vec {P}' \mid}{\mid \vec {P} \mid}$ increases very
rapidly and decreases moderately. The peak of
$\frac {1}{s}\frac {\mid \vec {P}' \mid}{\mid \vec {P} \mid}$ locates in the
slowly-changing region of $\int_0^\pi d\theta \mid {\cal M}_{\rm fi}(s,t) 
\mid^2\sin \theta$. Therefore, the peak of any curve
 in Figs. 2-8 near the threshold energy corresponds to
the peak of $\frac {1}{s}\frac {\mid \vec {P}' \mid}{\mid \vec {P} \mid}$.
In the present work we deal with meson-meson nonresonant reactions at low
 meson energies. Generally, only a small number of partial waves contribute to
low-energy reactions \cite{joachain}. Around the peak near the threshold
 energy, $s$ waves of the scattering constituents (quarks and/or antiquarks)
 contribute to the cross section. While $\sqrt s$ is away from the threshold
 energy, $p$ waves make a contribution to the cross section if a new peak is
 obvious, and $d$ waves do if one more new peak is obvious.

Now we pay attention to changes of peak cross sections with respect to
temperature in Figs. 2-8. The peak cross sections of
$\pi\pi \to \rho\rho$ for $I=2$, $KK \to K^* K^*$ for $I=1$,
$KK^* \to K^* K^*$ for $I=1$, $\pi K \to \rho K^*$ for $I=3/2$, and
$\pi K^* \to \rho K^*$ for $I=3/2$ have the same behavior: each decreases from
$T/T_{\rm c}=0$ to 0.85 and increases from $T/T_{\rm c}=0.85$ to 0.95.
 While the temperature increases from zero, the long-distance part of the
 potential gradually becomes a distance-independent and temperature-dependent
 value, confinement becomes weaker and weaker, the Schr\"odinger
 equation produces increasing meson radii, and mesonic bound states become
 looser and looser. The weakening confinement with increasing temperature
 makes it more difficult to combine final quarks and antiquarks into mesons
 through quark rearrangement in
 $q_1 \bar {q}_1 + q_2 \bar {q}_2 \to q_1 \bar {q}_2 + q_2 \bar {q}_1$, and
thus reduces cross sections. From $T/T_{\rm c}=0$ to 0.85 the amount of 
peak cross
section increased by the slowly increasing radii of initial $\pi$ and/or $K$
mesons can not balance the amount reduced by the weakening confinement,
and peak cross sections thus go down. From $T/T_{\rm c}=0.85$ to 0.95 the
amount of peak cross section increased by the rapidly increasing radii of
 initial $\pi$ and/or $K$ mesons overcomes the amount reduced by the weakening
 confinement, and peak cross sections thus go up.
 Similarly, we can understand that the peak cross section of
 $\rho K \to \rho K^*$ for $I=3/2$ decreases from $T/T_{\rm c}=0$ to 0.75
 and increases from $T/T_{\rm c}=0.75$ to 0.95 in Fig. 7.

In the above six reactions $\pi$ and $K$ appear only in the initial states. 
Unlike these reactions the reaction $\pi K^* \to \rho K$ for $I=3/2$ has
a pion in the initial state and a kaon in the final state. The slowly
increasing $K$ radius from $T=0$ makes the kaon formation less affected
by the weakening confinement. We thus get an increase in peak cross section
from $T/T_{\rm c}=0$ to 0.65. Depending on the amount of peak cross section
 increased by
the increasing $\pi$ radius and the amount reduced by the weakening 
confinement, there are a decrease in peak cross section from 
 $T/T_{\rm c}=0.65$ to 0.9 and an increase from $T/T_{\rm c}=0.9$ to 0.95.

\vspace{0.5cm}
\leftline{\bf 4. Summary}
\vspace{0.5cm}

We have established the quark potential which is given by perturbative QCD with
loop corrections at the short distance, becomes distance-independent and 
temperature-dependent at the long distance, and has the 
one-gluon-exchange spin-spin interaction smeared. The Schr\"odinger equation 
with the potential yields different temperature-dependent masses and different
quark-antiquark relative-motion wave functions for different mesons.
The potential, masses and wave functions in the transition amplitude 
developed from the
quark-interchange mechanism give temperature-dependent unpolarized cross
sections for the nonresonant reactions: $\pi\pi \to \rho\rho$ for $I=2$, 
$KK \to K^* K^*$ for $I=1$, $KK^* \to K^*K^*$ for $I=1$, 
$\pi K \to \rho K^*$ for $I=3/2$, $\pi K^* \to \rho K^*$ for $I=3/2$, 
$\rho K \to \rho K^*$ for $I=3/2$, and $\pi K^* \to \rho K$ for $I=3/2$.
The numerical cross sections are parametrized for convenient applications in 
hadronic matter in future. Every reaction takes a rise in peak cross sections
while the temperature approaches the critical temperature.

\vspace{0.5cm}
\leftline{ACKNOWLEDGEMENTS}
\vspace{0.5cm}
This work was supported by the National Natural Science Foundation of China
under Grant No. 11175111.

\newpage
\begin{figure}[htbp]
  \centering
    \includegraphics[scale=0.8]{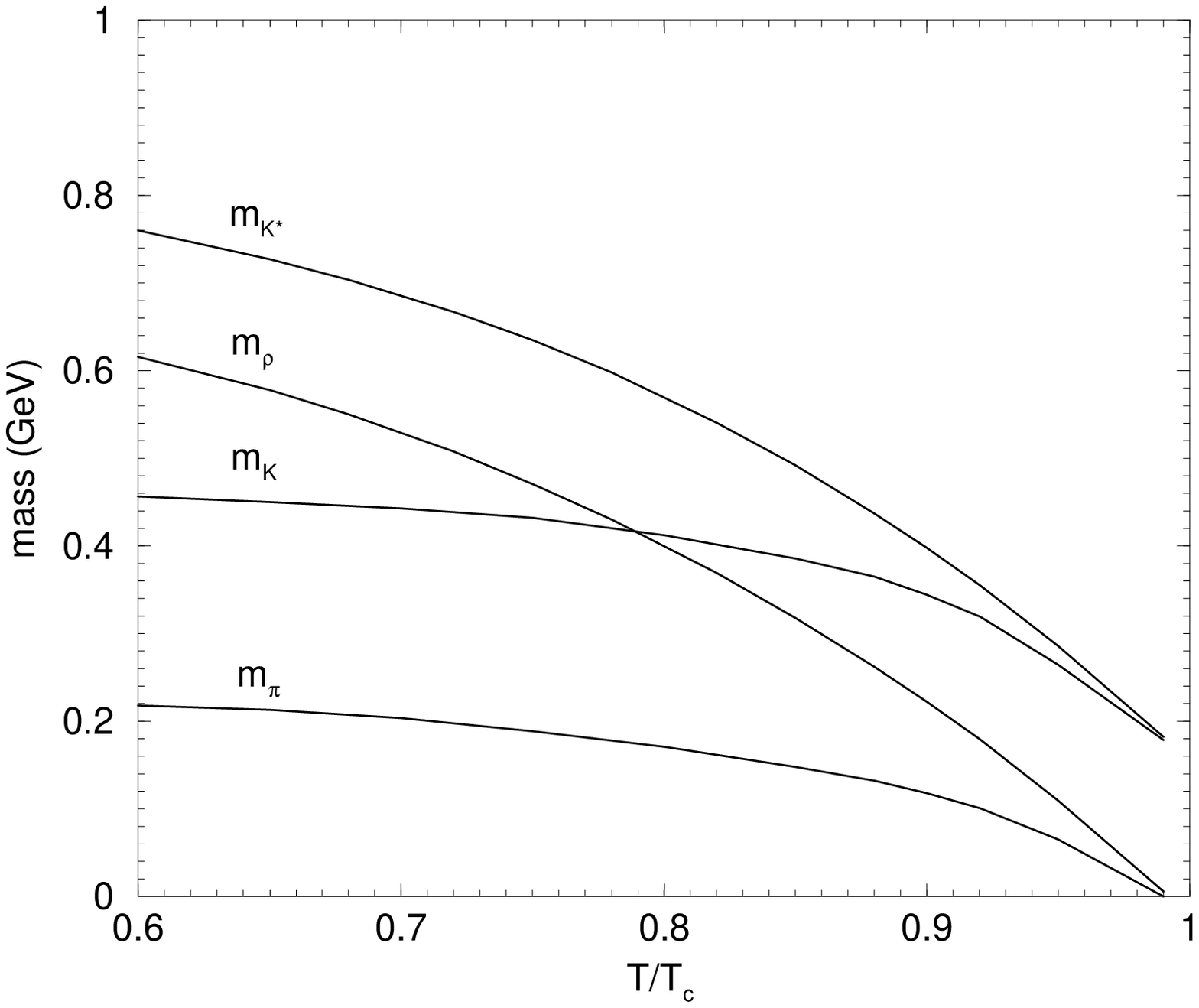}%
\caption{Meson masses as functions of $T/T_{\rm c}$.}
\label{fig1}
\end{figure}

\newpage
\begin{figure}[htbp]
  \centering
    \includegraphics[scale=0.8]{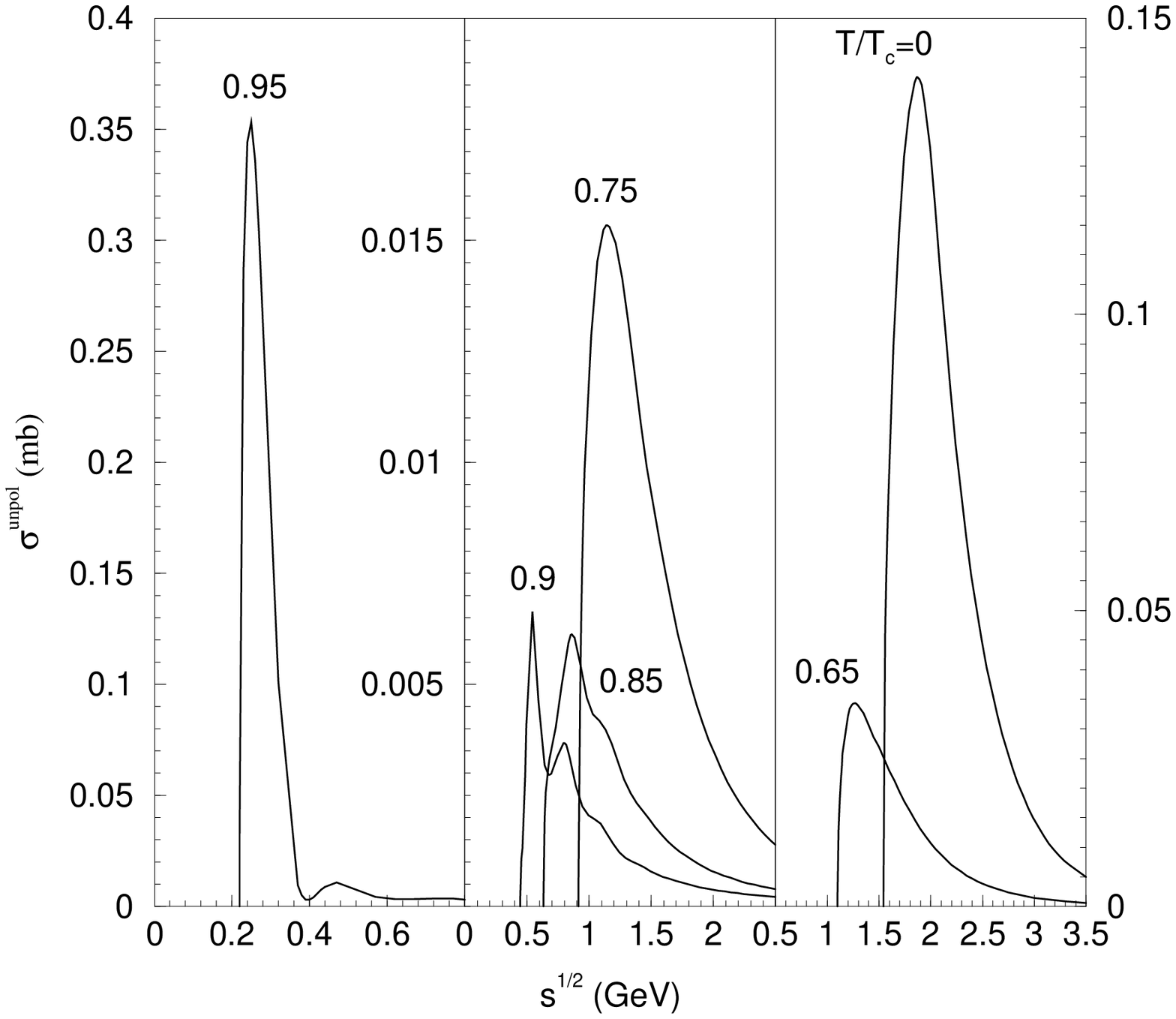}%
\caption{Cross sections for $\pi\pi\to\rho\rho$ for $I=2$ 
at various temperatures.}
\label{fig2}
\end{figure}

\newpage
\begin{figure}[htbp]
  \centering
    \includegraphics[scale=0.8]{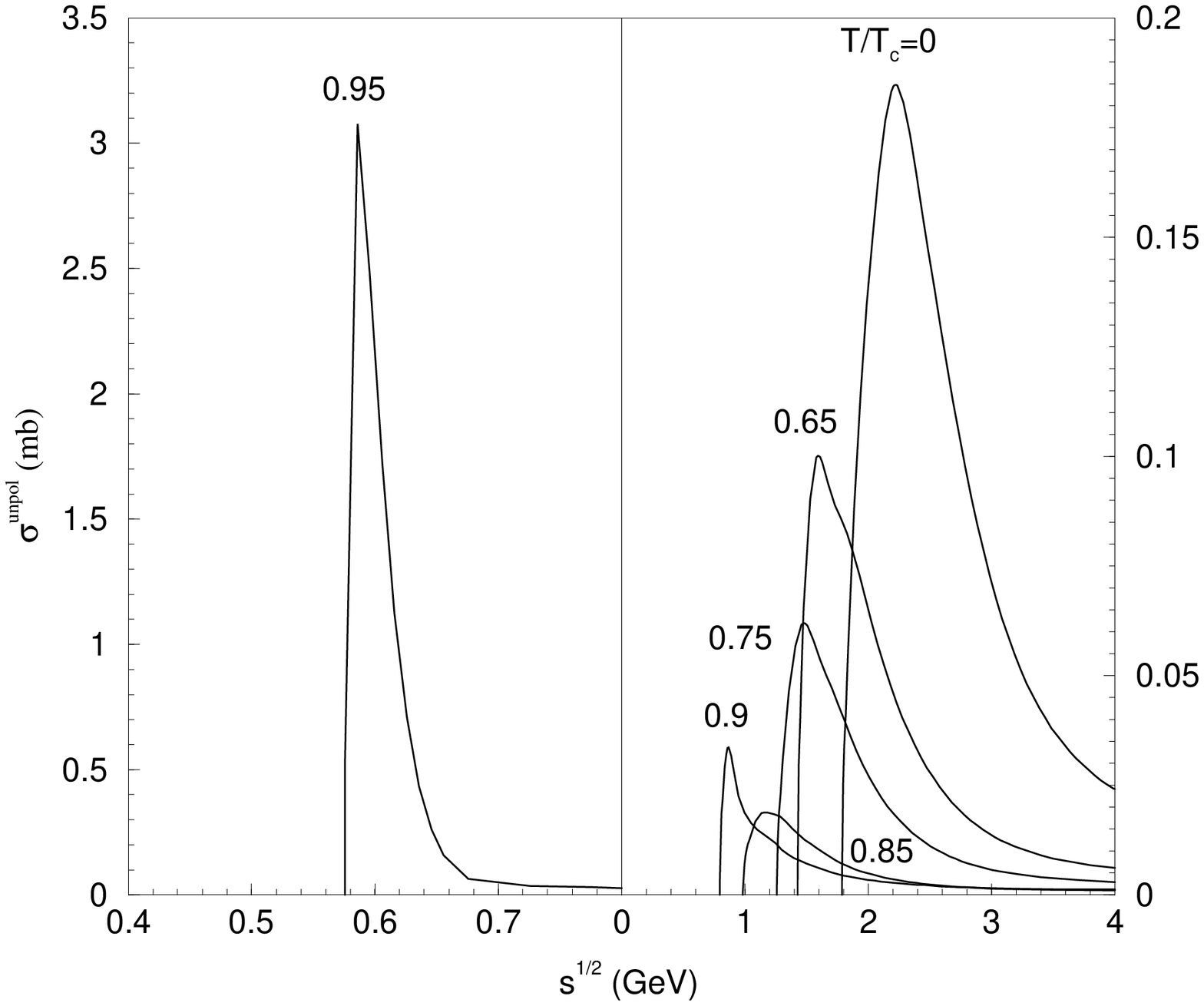}%
\caption{Cross sections for $KK \to K^* K^*$ for $I=1$
at various temperatures.}
\label{fig3}
\end{figure}

\newpage
\begin{figure}[htbp]
  \centering
    \includegraphics[scale=0.8]{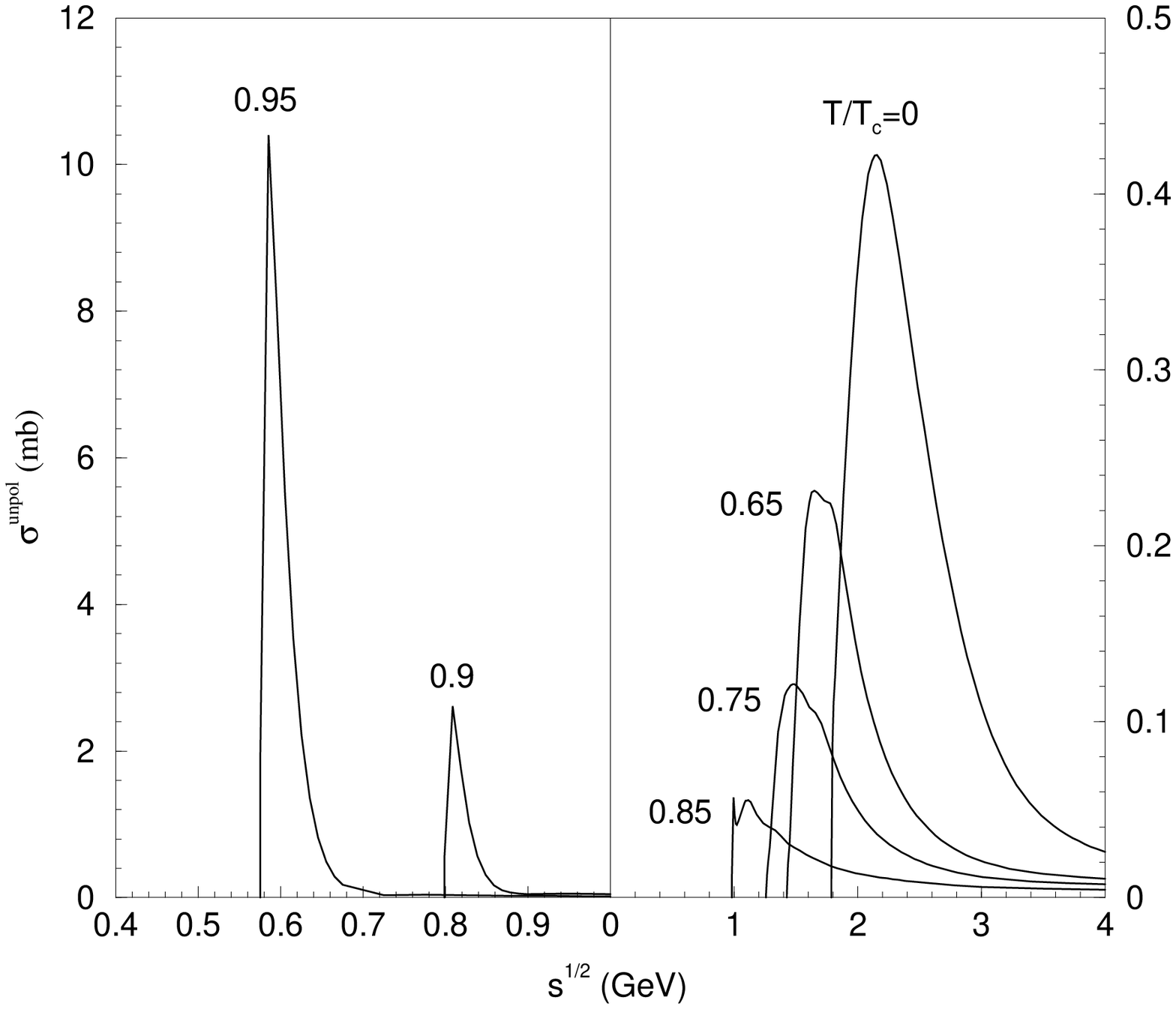}%
\caption{Cross sections for $KK^* \to K^* K^*$ for $I=1$
at various temperatures.}
\label{fig4}
\end{figure}

\newpage
\begin{figure}[htbp]
  \centering
    \includegraphics[scale=0.8]{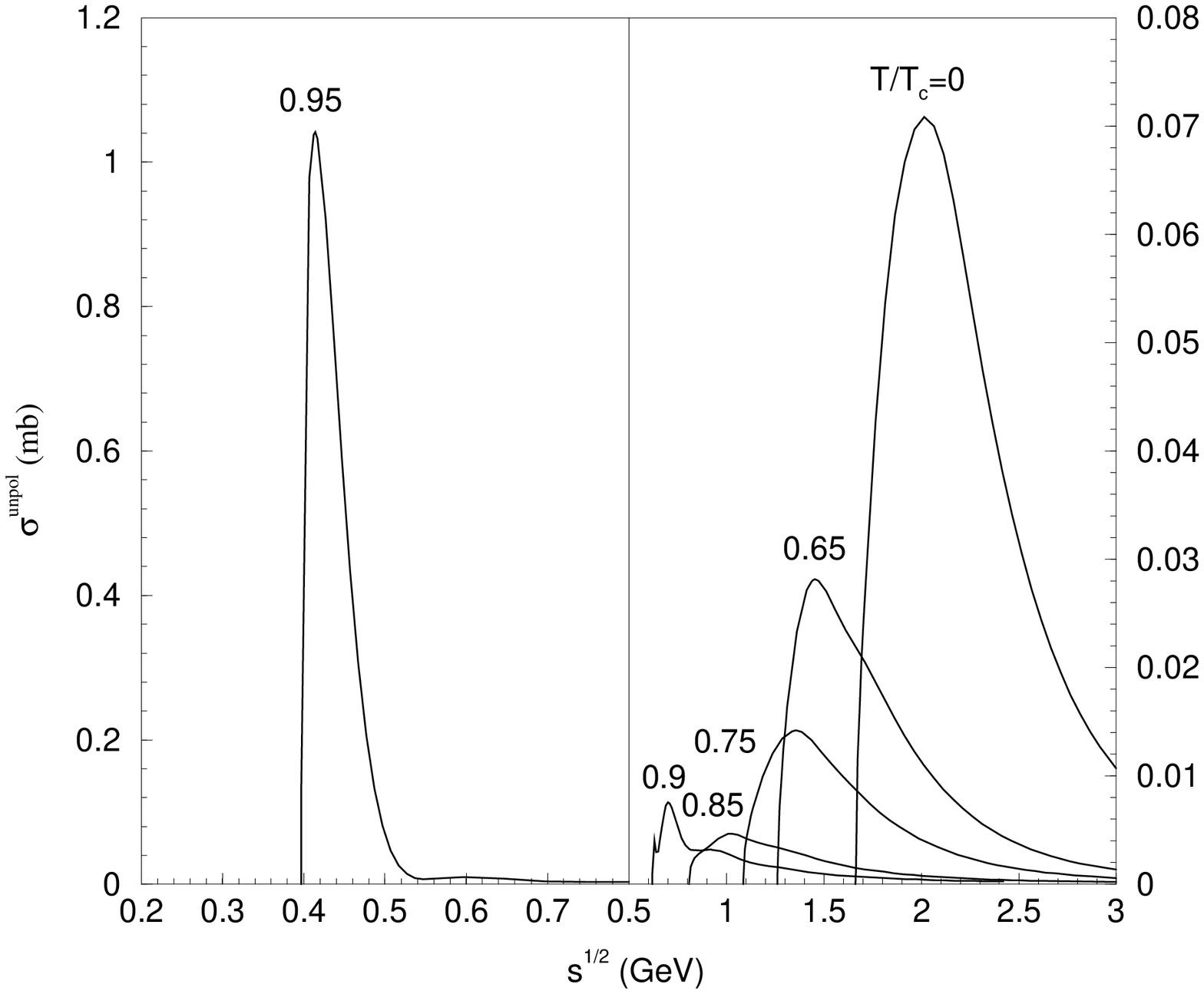}%
\caption{Cross sections for $\pi K \to \rho K^*$ for $I=3/2$
at various temperatures.}
\label{fig5}
\end{figure}

\newpage
\begin{figure}[htbp]
  \centering
    \includegraphics[scale=0.8]{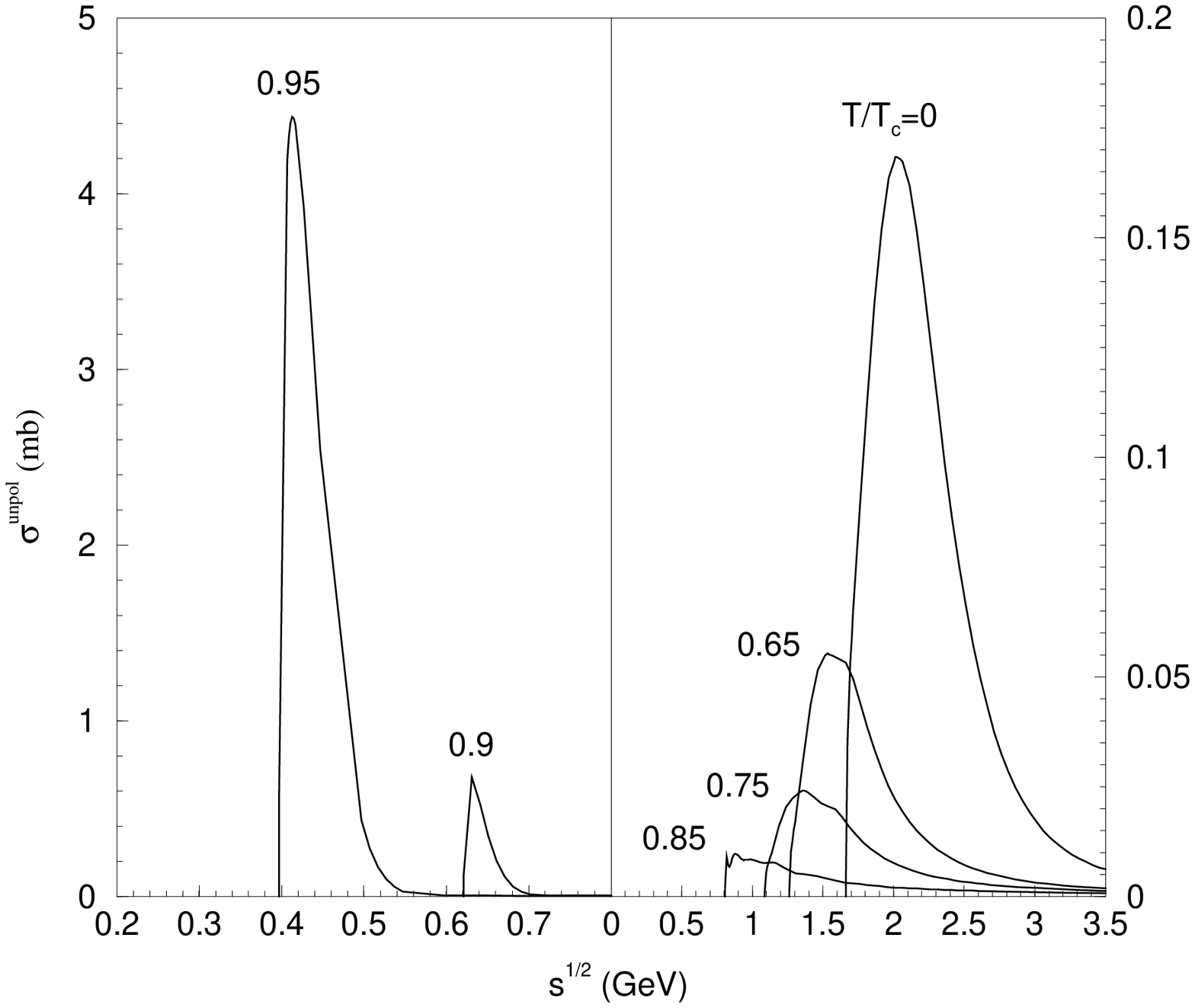}%
\caption{Cross sections for $\pi K^* \to \rho K^*$ for $I=3/2$
at various temperatures.}
\label{fig6}
\end{figure}

\newpage
\begin{figure}[htbp]
  \centering
    \includegraphics[scale=0.8]{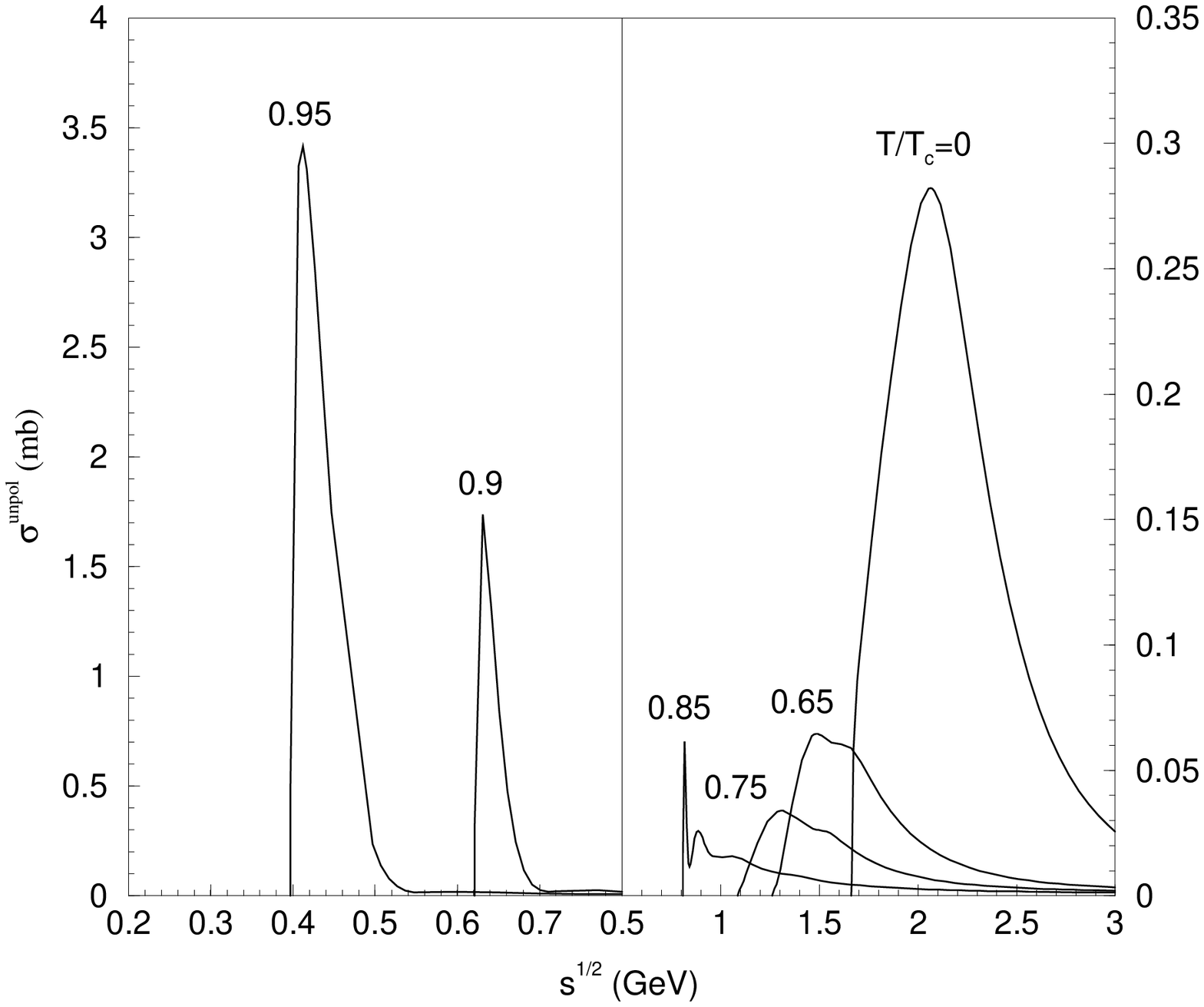}%
\caption{Cross sections for $\rho K \to \rho K^*$ for $I=3/2$
at various temperatures.}
\label{fig7}
\end{figure}

\newpage
\begin{figure}[htbp]
  \centering
    \includegraphics[scale=0.8]{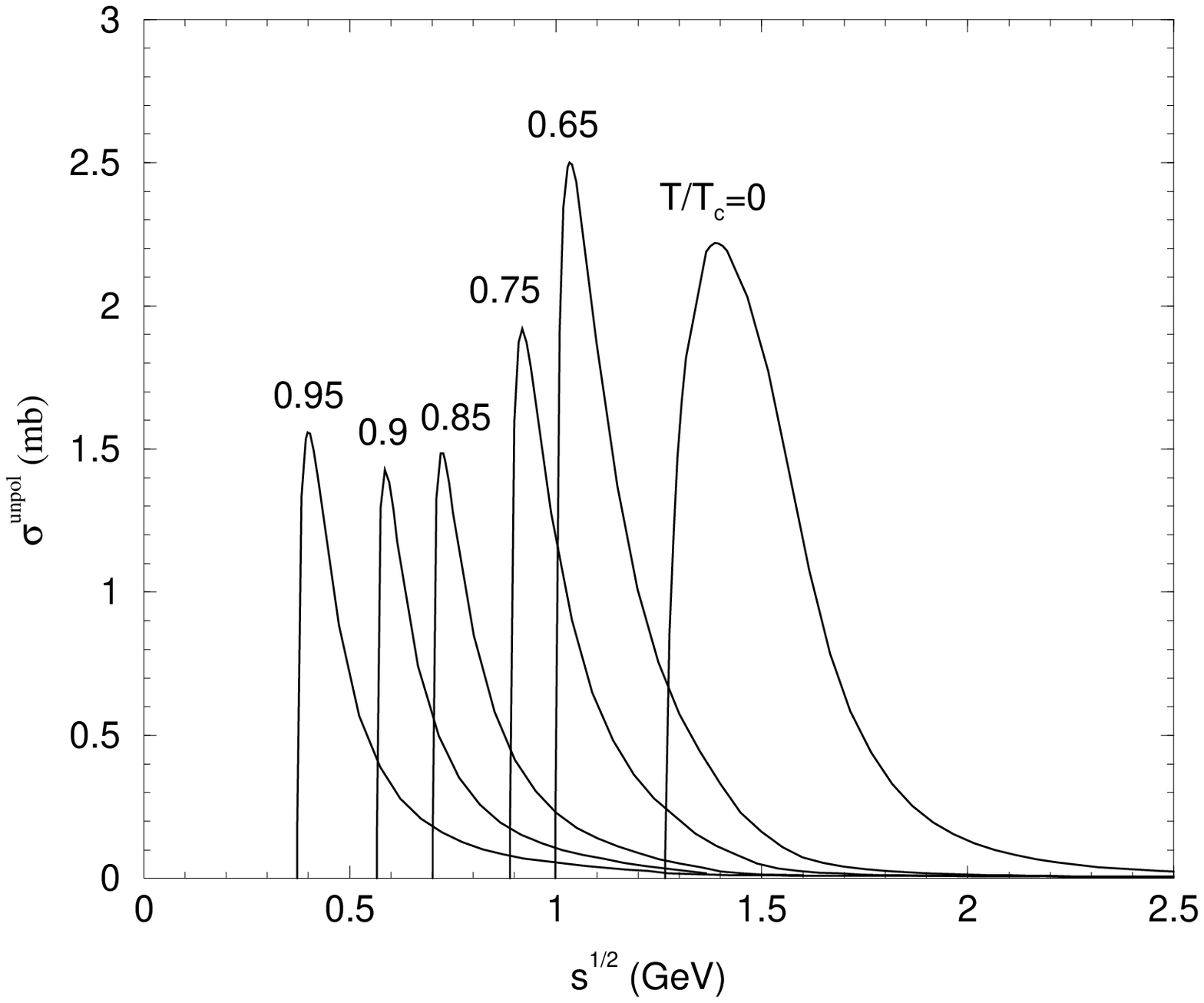}%
\caption{Cross sections for $\pi K^{*}\to \rho K$ for $I=3/2$
at various temperatures.}
\label{fig8}
\end{figure}

\newpage
\begin{table*}
\caption{\label{table:I}Values of the parameters in Eq. (15). $a_1$
and $a_2$ are in
units of millibarns; $b_1$, $b_2$, $d_0$, and $\sqrt{s_{\rm z}}$ are in units
of GeV; $c_1$ and $c_2$ are dimensionless. }
\begin{tabular}{cccccccccc}
\hline
Reactions & $T/T_{\rm c} $ & $a_1$ & $b_1$ & $c_1$ & $a_2$ & $b_2$ & $c_2$ &
$d_0$ & $\sqrt{s_{\rm z}}$ \\
\hline
 $\pi\pi\to\rho\rho$
  &  0     & 0.09  & 0.17  & 0.56  & 0.06  & 0.44  & 1.91  & 0.33 &4.272 \\
  &  0.65  & 0.022 & 0.25  & 0.62  & 0.014 & 0.1   & 0.46  & 0.17 &3.694 \\
  &  0.75  & 0.01  & 0.19  & 0.45  & 0.005 & 0.24  & 0.91  & 0.22 &3.221 \\
  &  0.85  & 0.002 & 0.1   & 0.4   & 0.004 & 0.28  & 0.9   & 0.23 &3.178 \\
  &  0.9   & 0.0036& 0.09  & 6.51  & 0.0032& 0.33  & 0.42  & 0.1 &3.113 \\
  &  0.95  & 0.192 & 0.01  & 0.49  & 0.249 & 0.04  & 1.64  & 0.03 &0.77 \\
  \hline
 $KK\to K^ {*} K^{*}$
  &  0     & 0.12  & 0.44  & 1.43  & 0.06  & 0.32  & 0.41  & 0.44 &5.21 \\
  &  0.65  & 0.06  & 0.18  & 0.77  & 0.04  & 0.28  & 0.38  & 0.17 &5.529 \\
  &  0.75  & 0.03  & 0.23  & 1.3   & 0.03  & 0.25  & 0.45  & 0.22 &5.73 \\
  &  0.85  & 0.012 & 0.12  & 0.45  & 0.008 & 0.36  & 0.94  & 0.18 &6.849 \\
  &  0.9   & 0.018 & 0.057 & 0.88  & 0.014 & 0.14  & 0.26  & 0.07 &8.208 \\
  &  0.95  & 0.2   & 0.0003 & 3.8  & 3.1   & 0.009 & 0.52  & 0.01 &0.777 \\
  \hline
 $KK^{*}\to K^* K^{*}$
  &  0     & 0.4   & 0.34  & 0.99  & 0.05  & 0.01  & 0.45  & 0.37 &5.005 \\
  &  0.65  & 0.19  & 0.25  & 1.69  & 0.05  & 0.44  & 0.44  & 0.22 &5.997 \\
  &  0.75  & 0.1   & 0.23  & 1.55  & 0.03  & 0.47  & 0.41  & 0.22 &6.782 \\
  &  0.85  & 0.039 & 0.008 & 0.3   & 0.047 & 0.18  & 0.9   & 0.13 &7.671 \\
  &  0.9   & 2.2   & 0.005 & 0.48  & 0.8   & 0.015 & 1.13  & 0.01 &1.207 \\
  &  0.95  & 8.4   & 0.009 & 0.59  & 2     & 0.007 & 0.26  & 0.01 &0.701 \\
\hline
\end{tabular}
\end{table*}

\begin{table*}
\caption{\label{table:II}The same as Table 1. }
\begin{tabular}{cccccccccc}
  \hline
  Reactions & $T/T_{\rm c} $ & $a_1$ & $b_1$ & $c_1$ & $a_2$ & $b_2$ & $c_2$ &
  $d_0$ & $\sqrt{s_{\rm z}} $ \\
  \hline
  $\pi K\to\rho K^{*}$
  &  0     & 0.038  & 0.345 & 1.95  & 0.034   & 0.24  & 0.49  & 0.35 &4.008 \\
  &  0.65  & 0.018  & 0.2   & 0.47  & 0.01    & 0.21  & 1.5   & 0.19 &3.836 \\
  &  0.75  & 0.0091 & 0.21  & 0.5   & 0.005   & 0.288 & 2.27  & 0.27 &3.878 \\
  &  0.85  & 0.0009 & 0.28  & 3.5   & 0.00373 & 0.184 & 0.46  & 0.22 &4.112 \\
  &  0.9   & 0.0032 & 0.005 & 0.5   & 0.0061  & 0.081 & 2     & 0.08 &3.031 \\
  &  0.95  & 0.91   & 0.011 & 0.53  & 0.32    & 0.037 & 2.59  & 0.017 &0.534 \\
  \hline
  $\pi K^{*}\to \rho K^{*}$
   & 0     & 0.16   & 0.38   & 2     & 0.05   & 0.05  & 0.49  & 0.36 &4.032 \\
   & 0.65  & 0.034  & 0.31   & 2.57  & 0.024  & 0.21  & 0.44  & 0.27 &4.112 \\
   & 0.75  & 0.0105 & 0.146  & 0.41  & 0.0159 & 0.38  & 2     & 0.28 &4.352 \\
   & 0.85  & 0.0038 & 0.0039 & 0.59  & 0.0092 & 0.134 & 0.32  & 0.07 &4.819 \\
   & 0.9   & 0.45   & 0.002  & 0.72  & 0.61   & 0.012 & 0.96  & 0.01 &0.805 \\
   & 0.95  & 0.3    & 0.001  & 0.006 & 4.2    & 0.017 & 0.68  & 0.016 &0.542 \\
  \hline
  $\rho K\to \rho K^{*}$
   & 0     & 0.18   & 0.42   & 2.9   & 0.124  & 0.151 & 0.45  & 0.4  &3.859 \\
   & 0.65  & 0.0045 & 0.11   & 0.39  & 0.056  & 0.304 & 1.77  & 0.23 &4.345 \\
   & 0.75  & 0.0259 & 0.239  & 1.74  & 0.0084 & 0.802 & 0.44  & 0.23 &4.858 \\
   & 0.85  & 0.05   & 0.0029 & 0.31  & 0.022  & 0.097 & 1.41  & 0.08 &5.728 \\
   & 0.9   & 1.3    & 0.007  & 0.44  & 0.5    & 0.013 & 1.4   & 0.01 &0.811 \\
   & 0.95  & 2.2    & 0.002  & 1.32  & 3.4    & 0.015 & 0.58  & 0.015 &0.532 \\
  \hline
  $\pi K^{*}\to \rho K$
   & 0     & 1.3   & 0.05  & 0.54  & 1.49  & 0.188 & 1.96  & 0.12 &2.549 \\
   & 0.65  & 0.97  & 0.1   & 0.66  & 1.8   & 0.027 & 0.49  & 0.035 &1.812 \\
   & 0.75  & 1.17  & 0.02  & 0.493 & 0.94  & 0.071 & 0.545 & 0.03 &1.657 \\
   & 0.85  & 0.9   & 0.018 & 0.53  & 0.7   & 0.056 & 0.44  & 0.02 &1.494 \\
   & 0.9   & 0.47  & 0.061 & 0.38  & 1.02  & 0.02  & 0.52  & 0.02 &1.386 \\
   & 0.95  & 1.02  & 0.022 & 0.55  & 0.6   & 0.054 & 0.39  & 0.025 &1.314  \\
  \hline
\end{tabular}
\end{table*}

\end{document}